\documentclass[amsmath,amssymb,12pt,superscriptaddress,nofootinbib]{revtex4}

\usepackage{graphicx}
\usepackage{dcolumn}
\usepackage{bm}
\usepackage{color}
\usepackage{aas_macros}

\newcommand{\dd}{\mathrm{d}}

\newcommand{\del}{\partial}
\newcommand{\ee}{{\rm e}}

\newcommand{\cm}{{\rm cm}}

\newcommand{\dl}{D_{\rm L}}
\newcommand{\ds}{D_{\rm S}}
\newcommand{\dls}{D_{\rm LS}}

\definecolor{DarkBlue}{rgb}{0,0,0.7} 

\definecolor{DarkRed}{rgb}{0.65,0,0}

\begin{document}
\baselineskip5.5mm
\thispagestyle{empty}

{\baselineskip0pt
\leftline{\baselineskip14pt\sl\vbox to0pt{
               \hbox{\it Yukawa Institute for Theoretical Physics} 
              \hbox{\it Kyoto University}
              \medskip
               \hbox{\it Department of Physics} 
              \hbox{\it Rikkyo University}
               \vss}}
\rightline{\baselineskip16pt\rm\vbox to20pt{
            \hbox{YITP-13-15}
            \hbox{RUP-13-3}
\vss}}%
}

\author{Chul-Moon Yoo}\email{yoo@yukawa.kyoto-u.ac.jp}
\affiliation{
Yukawa Institute for Theoretical Physics, Kyoto University
Kyoto 606-8502, Japan
}

\author{Tomohiro Harada}%\email{}
\affiliation{
Department of Physics, Rikkyo University, Tokyo 171-8501, Japan}

\author{Naoki Tsukamoto}%\email{}
\affiliation{
Department of Physics, Rikkyo University, Tokyo 171-8501, Japan}

\vskip3cm
\title{Wave Effect in Gravitational Lensing by the Ellis Wormhole}

\begin{abstract}
We propose the use of modulated spectra 
of astronomical sources 
due to gravitational lensing to probe
Ellis wormholes. 
The modulation factor 
due to gravitational lensing by 
the Ellis wormhole is calculated. 
Within the geometrical optics approximation, the normal point 
mass lens and the Ellis wormhole are indistinguishable unless 
we know the source's unlensed luminosity. 
This degeneracy is 
resolved with the significant wave effect in the low frequency 
domain if we take the deviation from the geometrical optics 
into account. 
We can roughly estimate the upper bound for the number density 
of Ellis wormholes as 
$n\lesssim 10^{-9}\mbox{AU}^{-3}$
with 
throat radius $a\sim1\cm$ from 
the existing femto-lensing analysis for compact objects. 
\end{abstract}

\maketitle
\pagebreak
%%%%%%%%%%%%%%%%%%%%%%%%%%%%%%%%%%%%%%%%%%%%%%%%%%%%%%%%%%%%%%%%
\section{introduction}
\label{sec:intro}
%%%%%%%%%%%%%%%%%%%%%%%%%%%%%%%%%%%%%%%%%%%%%%%%%%%%%%%%%%%%%%%%
In a variety of cosmological models based on fundamental theory,
exotic astrophysical objects which have not been observed are often 
predicted. 
Conversely, an observational evidence for an exotic object would 
stimulate creative theoretical discussions. 
Probing these exotic objects and detecting them will give us 
significant progress of research in fundamental physics. 
Even if we cannot
detect it, 
giving a constraint on the abundance of the exotic objects is 
one of powerful means of investigating 
the nature of our universe.

Generally, the interaction between 
such unobserved exotic objects and well known matters 
is very weak or 
%and 
not well established. 
Thus, only the gravitational interaction 
would cause reliable observational phenomena. 
One of the most direct measurements of gravitational 
effects of an exotic object is gravitational lensing. 
For instance, 
massive compact halo objects 
are probed by using 
micro-lensing\cite{Alcock:2000ph,Tisserand:2006zx,Wyrzykowski:2009ep}. 
Cosmic strings are also targets for probing by using 
gravitational lensing 
phenomena\cite{Huterer:2003ze,Oguri:2005dt,Mack:2007ae,2008MNRAS.384..161K,Christiansen:2010zi,Tuntsov:2010fu,Pshirkov:2009vb,Yamauchi:2011cu,Yamauchi:2012bc}. 
In this paper, we propose a way to probe
Ellis wormholes\cite{Ellis:1973yv} by using lensed spectra of astronomical sources.

The Ellis wormhole was first 
introduced by Ellis as a spherically symmetric 
solution of Einstein equations with a ghost massless scalar field. 
The dynamical stability of the Ellis wormhole is discussed in Ref.~\cite{Shinkai:2002gv} 
and the possible source to support the Ellis geometry was proposed in 
Ref.~\cite{Das:2005un}. 
Gravitational lensing by the Ellis wormhole was studied in Refs.~\cite{1984GReGr..16..111C,1984IJTP...23..335C} and recently revisited by several 
authors\cite{Nakajima:2012pu,Tsukamoto:2012zz}. 
So far, it has been suggested that 
Ellis wormholes can be probed by using 
light curves of gamma-ray bursts~\cite{Torres:1998xd}, 
micro-lensing~\cite{Safonova:2001vz,Bogdanov:2008zy,Abe:2010ap}
(see also Refs.~\cite{Asada:2011ap,Kitamura:2012zy}) and imaging 
observations~\cite{Toki:2011zu,Tsukamoto:2012xs}, 
while our proposal is the use of spectroscopic observations 
to probe Ellis wormholes. 

In order to fully investigate the lensed spectrum of a point source, 
the wave effect in gravitational lensing must be 
taken into account. 
The wave effect for the point mass lens  is 
%is 
discussed in 
Refs.~\cite{Deguchi:1986zz,1992grle.book.....S}. 
Wave effects in gravitational lensing by the 
rotating massive object\cite{Baraldo:1999ny}, 
binary system\cite{Mehrabi:2012dy}, 
%the 
singular isothermal sphere\cite{Takahashi:2003ix} 
and the cosmic string\cite{Suyama:2005ez,Yoo:2012dn} 
have been considered. 
In Sec.~\ref{waveform}, we calculate the amplification factor 
of gravitational lensing by the Ellis wormhole 
taking the wave effect into account. 
The geometrical optics limit is analytically 
presented in Sec.~\ref{GOA}. 
The difference in the amplification factor between 
the point mass lens and the Ellis wormhole lens is 
discussed in Sec.~\ref{obs} based on observables. 
In Sec.~\ref{obsconst}, possible observations to 
probe Ellis wormholes are listed. 
Sec.~\ref{summary} is devoted to a summary. 

In this paper, we use the geometrized units in which 
the speed of light and 
Newton's 
gravitational constant are 
both unity.
%one, respectively.

%%%%%%%%%%%%%%%%%%%%%%%%%%%%%%%%%%%%%%%%%%%%%%%%%%%%%%%%%%%%%%%%
\section{A Derivation of the Lensed Wave Form}
\label{waveform}
%%%%%%%%%%%%%%%%%%%%%%%%%%%%%%%%%%%%%%%%%%%%%%%%%%%%%%%%%%%%%%%
The line element in the Ellis wormhole spacetime 
can be written by the following 
isotropic form: 
\begin{equation}
\dd s^2=-\dd t^2+\left(1+\frac{a^2}{R^2}\right)^2\left(\dd R^2
+R^2\dd \Omega^2\right), 
\end{equation}
where $R=a$ corresponds to the throat and we simply call 
$a$ the throat radius in this paper.%
\footnote{Since the throat surface area is given by $16\pi a^2$, 
our definition of the throat radius is 
half the areal radius of the throat. }

\begin{figure}[htbp]
\includegraphics[scale=1]{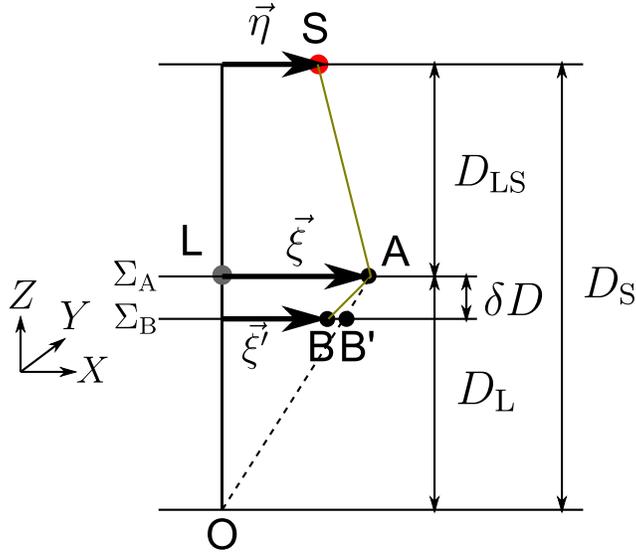}
\caption{Lens system with thin lens approximation. 
S, L, and O represent the source, lens, and observer positions, 
respectively. 
The %line 
path SAB is a ray trajectory which is specified with 
the vector $\vec \xi$ on the lens plane $\Sigma_{\rm A}$. 
${\rm B'}$ is the intersection of the line AO and the plane $\Sigma_{\rm B}$. 
$\vec \xi'$ is the position vector of the point B on the plane $\Sigma_{\rm B}$.
}
\label{thinlens}
\end{figure}

Assuming the thin lens approximation is valid, 
we consider the wormhole lens system shown in Fig.~\ref{thinlens}. 
%%%%%%%%%%%%%%%%%%%%%%%%%%%<<start figure>>%%%%%%%%%%%%%%%%%%%%%%%%%%
%%%%%%%%%%%%%%%%%%%%%%%%%%%%<<end figure>>%%%%%%%%%%%%%%%%%%%%%%%%%%%
We use the position vector $\vec X=(X,Y,Z)$ in the 
flat space. 
Then, the coordinate $R$ is given by  
$R=|\vec X-\vec X_{\rm L}|$, where $|\vec X_{\rm L}|$ is 
the lens position. 
We set $Z$-axis as the perpendicular direction to the lens plane and the 
source plane. 
$\ds$, $\dl$, and $\dls$ denote 
the distances from the observer plane to the source plane, from 
the observer plane to the lens plane, and from the lens plane to the source plane, 
respectively. 

In the geometrical optics limit, 
we consider light rays emanated from the source. 
The vector $\vec \xi$ on the lens plane $\Sigma_{\rm A}$ in Fig.~\ref{thinlens}
specifies the light ray 
which is deflected once at $\vec X=\vec X_{\rm L}+\vec \xi$. 
Since $\xi:=|\vec \xi|$ can be regarded as the 
closest approach of the light ray, as is shown in Ref.~\cite{1984GReGr..16..111C}, 
the deflection angle $\alpha$ is given by 
\begin{equation}
\alpha(\xi)=\pi\left(\frac{a}{\xi}\right)^2+\mathcal O\left(\frac{a}{\xi}\right)^4. 
\end{equation}
As a result, the %The 
Einstein radius $\xi_0$ for the Ellis wormhole 
is given by 
\begin{equation}
\xi_0=\left(\frac{1}{4}\pi a^2D\right)^{1/3},  
\end{equation}
where 
\begin{equation}
D=\frac{4\dl\dls}{\ds}. 
\end{equation}

Since we are interested in the wave effect, that is, the deviation from 
the geometrical optics limit, we need to treat the wave equation rather than 
light rays. 
Neglecting the polarization effect, we consider the scalar wave equation with the frequency $\omega$.  
The wave equation for the monochromatic wave $\ee^{i\omega t}\phi(\vec X)$ 
is given by 
\begin{equation}
\omega^2\phi+\left(1+\frac{a^2}{R^2}\right)^{-3}\del_i
\left[\left(1+\frac{a^2}{R^2}\right)\delta^{ij}\del_j\phi\right]
=
-4\pi A ~\delta(\vec X-\vec X_{\rm S}), 
\end{equation}
where $\vec X_{\rm S}$ is the position vector of the point source and 
$A$ in the source term is a constant which specifies the amplitude. 
Without the wormhole, we obtain 
the wave form $\bar \phi_{\rm O}$ at the observer O  
as follows:
\begin{equation}
\bar \phi_{\rm O}=\frac{A}{\sqrt{\ds^2+\eta^2}}
\exp\left[i\omega \sqrt{\ds^2+\eta^2}\right]
=
\frac{A}{\ds}
\exp\left[i\omega \ds\left(1+\frac{\eta^2}{2\ds^2}
+\mathcal O\left(\left(\frac{\eta}{\ds}\right)^4\right)\right)\right], 
\end{equation}
where $\eta=|\vec \eta|$ and $\vec \eta=(X_{\rm S}-X_{\rm L},Y_{\rm S}-Y_{\rm L},0)$ 
and we consider the case $\eta\ll \ds$ in this paper. %

Our assumptions to calculate the wave form at O 
are  
summarized as follows(see Ref.~\cite{1992grle.book.....S}): 
\begin{itemize}
\item[(a)] The geometrical optics approximation is 
valid between the source plane and the plane $\Sigma_{\rm B}$ 
in Fig.~\ref{thinlens}. 
\item[(b)] Thin lens approximation is valid and  
a ray from the source is deflected once on the lens plane $\Sigma_{\rm A}$. 
\item[(c)] 
Assuming $\ds\sim\dl\sim\dls\sim D$, 
we use a non-dimensional parameter $\epsilon$ 
defined by $\epsilon:=\xi_0/D$,
which gives the typical scale of the deflection angle. 
Then, 
we assume $1/(\omega D) \ll \epsilon \ll 1$ and $\eta/D= \mathcal O(\epsilon)$.
\item[(d)] On the plane $\Sigma_{\rm B}$, the gravitational potential of the 
lens object is negligible and $\delta D/D= \mathcal O(\epsilon)$, 
where $\delta D$ is the distance between the planes $\Sigma_{\rm A}$ and 
$\Sigma_{\rm B}$. 
\end{itemize}
On the assumptions made above, we calculate 
the wave form on the plane $\Sigma_{\rm B}$ 
up to the leading order for the amplitude and 
next leading terms for the phase part. 
Applying the Kirchhoff integral theorem between the plane $\Sigma_{\rm B}$ and 
the observer, we calculate the approximate wave form at O. 

The vector $\vec \xi$ specifies the light ray 
which is deflected once on the plane 
$\Sigma_{\rm A}$ at $\vec X=\vec X_{\rm L}+\vec \xi$ 
and reaches the plane $\Sigma_{\rm B}$. 
The deflection angle is fixed by $\vec \xi$ 
and the background geometry. 
The point A in Fig.~\ref{thinlens} denotes 
the deflected point.
We label 
the intersection of the deflected light ray and 
the plane $\Sigma_{\rm B}$ as B, while 
${\rm B'}$ in the Fig.~\ref{thinlens} denotes the intersection 
of the line AO and the plane $\Sigma_{\rm B}$. 
As will be mentioned at the end of this section, 
the dominant contribution 
to the wave form at O 
comes from rays which satisfy $\xi\sim \epsilon D$, 
where $\xi=|\vec \xi|$. 
Therefore we consider $\vec \xi/D$ as $\mathcal O(\epsilon)$ hereafter.

First, we consider the following ansatz for $\phi$ in the region between 
the source plane and the plane $\Sigma_{\rm B}$:
\begin{equation}
\phi=f(\vec X)~\ee^{iS(\vec X)}. 
\end{equation}
On the plane $\Sigma_{\rm B}$, the amplitude $f(\vec X)$ is given by 
\begin{equation}
\left.f(\vec X)\right|_{\Sigma_{\rm B}}
=\frac{A}{\dls}\left(1+\mathcal O(\epsilon)\right). 
\label{amplitude1}
\end{equation}
In the geometrical optics approximation, 
the phase $S(\vec X)$ satisfies the eikonal equation given by 
\begin{equation}
%-\omega^2+g^{ij}\del_iS\del_jS=0 
%\Leftrightarrow 
\delta^{ij}\del_i S\del_j S =\omega^2\left(1+\frac{a^2}{R^2}\right)^2. 
\end{equation}
At the point B, the phase based on the source position S is 
given by the following integral 
\begin{equation}
\left.S\right|_{\rm B}=\int^{\rm B}_{\rm S}
\frac{\dd x^i}{\dd l}\del_iSdl, 
\label{sourceint1}
\end{equation}
where we have introduced the optical path length $l$ defined as
\begin{equation}
\delta_{ij}\frac{\dd x^i}{\dd l}\frac{\dd x^j}{\dd l}=1. 
\end{equation}
Since, in the geometrical optics approximation, we find 
\begin{equation}
\del_j S=\omega\left(1+\frac{a^2}{R^2}\right)\delta_{ij}\frac{\dd x^i}{\dd l}, 
\end{equation}
the integral \eqref{sourceint1} is 
given by 
\begin{equation}
\left.S\right|_{\rm B}=
\omega\int^{\rm B}_{\rm S} dl+\omega a^2\int^{\rm B}_{\rm S}\frac{1}{R^2} dl. 
\label{sourceint2}
\end{equation}
After the calculations 
explicitly shown in Appendix~\ref{dereq}, 
we finally obtain the following expression: 
\begin{equation}
\left.S\right|_B\simeq\omega\left[
\ds\left(1+\frac{\eta^2}{2\ds^2}\right)+
\frac{\dl\ds}{2\dls}\left(\frac{\vec \xi}{\dl}-\frac{\vec \eta}{\ds}\right)^2
-r+\frac{\pi a^2}{\xi}\right]. 
\label{sourceint3}
\end{equation}
This expression for the phase and Eq.~\eqref{amplitude1} 
for the amplitude can be used for any value of $\vec \xi$, that is, 
we have obtained an approximate wave form on the plane $\Sigma_{\rm B}$.

Applying the Kirchhoff integral theorem\cite{1999prop.book.....B} and neglecting 
the contribution from the infinity, 
we express the wave form $\phi_{\rm O}$ at O by the following integral:
\begin{equation}
\phi_{\rm O}=-\frac{1}{4\pi}\int_{\Sigma_{\rm B}} \dd \xi^2 \left\{
\phi_{\rm B}\frac{\del}{\del Z}\left(\frac{\ee^{i\omega r}}{r}\right)
-\frac{\ee^{i\omega r}}{r}\frac{\del \phi_{\rm B}}{\del Z} \right\}, 
\end{equation}
where $\phi_{\rm B}$ is the waveform at B. 
Since we are interested in only the leading order of 
the amplitude, we obtain 
\begin{equation}
\phi_{\rm O}\simeq
-\frac{i\omega A}{2\pi \dl \dls}
\exp\left[i\omega\ds\left(1+\frac{\eta^2}{2\ds^2}\right)\right]
\int \dd \xi^2 
\exp\left[i\omega\left\{
+\frac{\dl\ds}{2\dls}\left(\frac{\vec\xi}{\dl}-\frac{\vec\eta}{\ds}\right)^2
+\frac{\pi a^2}{\xi}\right\}\right], 
\end{equation}
where we have used the following approximations:
\begin{eqnarray}
&&\frac{\del}{\del Z}\left(\frac{\ee^{i\omega r}}{r}\right)\simeq 
\frac{i\omega}{r}\ee^{i\omega r}\simeq \frac{i\omega}{\dl}\ee^{i\omega r}, 
\\
&&\frac{\del \phi_{\rm B}}{\del Z} 
\simeq \frac{-i\omega A}{\dls}\ee^{i\left.S\right|_B}. 
\end{eqnarray}

Defining the amplification factor $F$ by $F:=\phi_{\rm O}/\bar \phi_{\rm O}$, 
we obtain
\begin{equation}
F\simeq \frac{\omega d}{\pi i}\int \dd x^2
\exp\left[i\omega d\left\{(\vec x-\vec y)^2+\frac{2}{x}\right\}\right], 
\end{equation}
where 
\begin{eqnarray}
d&=&\frac{\xi_0^2\ds}{2\dls\dl}=\frac{2\xi_0^2}{D}=2\xi_0\epsilon,
\label{d4wh}\\
\vec x&=&\frac{\vec\xi}{\xi_0}, \\
\vec y&=&\frac{\vec \eta\dl}{\xi_0 \ds}. 
\end{eqnarray}
$d$ gives the optical path difference 
between the lensed trajectory and the unlensed one in the 
geometrical optics limit for $\vec\eta=0$. 

\begin{figure}[htbp]
\includegraphics[scale=1]{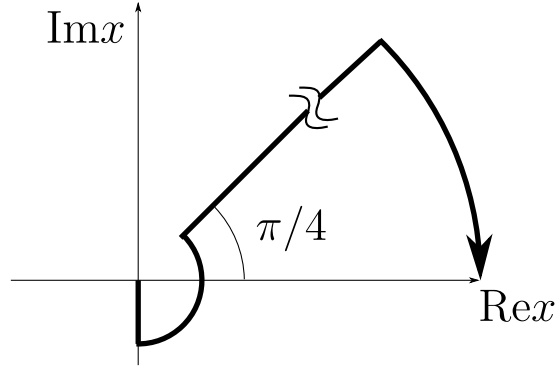}
\caption{The path of the integral \eqref{intx} taken in the numerical integration. 
}
\label{path}
\end{figure}

Introducing a polar coordinate, we rewrite this integral as
\begin{eqnarray}
F&\simeq&
\frac{\omega d}{\pi i}\ee^{i\omega d y^2}\int^\infty_0\dd x 
x\exp\left[i\omega d\left(x^2+\frac{2}{x}\right)\right]\int^{2\pi}_0\dd\varphi
\exp\left[-2i\omega d xy \cos\varphi\right]
\label{intxphi}
\\
&=&
-2i\omega d\ee^{i\omega dy^2}\int^\infty_0\dd x 
x\exp\left[i\omega d\left(x^2+\frac{2}{x}\right)\right]
J_0(2\omega dxy). 
\label{intx}
\end{eqnarray}
The integrand is divergent at the infinity on the real axis. 
This is caused by our approximation associated with $\epsilon$, 
and not real. 
If we write down the integrand in a 
precise form without 
any approximation, 
we do not have any divergence. 
Actually, in the 
precise form, the contribution from 
the integral in the region $x\gg 1$ is negligible due to 
the cancellation of the quasi-periodic integration. 
For the same reason, 
the contribution from the integration in the region 
$x\ll 1$ is also negligible. 
Since the approximate expression of Eq.~\eqref{intx} is valid in 
the region $x\ll1/\epsilon$, 
we can obtain a reliable result by 
neglecting the contribution from the region $x\gg 1$ in 
the integral \eqref{intx}. 
Practically, to make this integral finite, it is convenient to 
consider the analytic continuation to the complex plane 
and take the path of the integral 
as shown in Fig.~\ref{path}. 
Then, this integral can be numerically performed. 
%%%%%%%%%%%%%%%%%%%%%%%%%%%<<start figure>>%%%%%%%%%%%%%%%%%%%%%%%%%%
%%%%%%%%%%%%%%%%%%%%%%%%%%%%<<end figure>>%%%%%%%%%%%%%%%%%%%%%%%%%%%

%%%%%%%%%%%%%%%%%%%%%%%%%%%%%%%%%%%%%%%%%%%%%%%%%%%%%%%%%%%%%%%%
\section{Geometrical Optics Approximation}
\label{GOA}
%%%%%%%%%%%%%%%%%%%%%%%%%%%%%%%%%%%%%%%%%%%%%%%%%%%%%%%%%%%%%%%%

In this section, we derive an approximate 
form of the amplification factor $F$. 
In the expression \eqref{intxphi}, 
we apply the stationary phase approximation 
to the integral with respect to $x$. 
Then we obtain 
\begin{equation}
F\simeq\sqrt{\frac{\omega d}{\pi i}}\int^{2\pi}_0
\dd\varphi \frac{x_0}{\sqrt{1+2/x_0^3}}\exp\left[i\omega d h(x_0)\right], 
\label{geointx}
\end{equation}
where 
\begin{equation}
h(x)=x^2-2xy\cos\varphi+\frac{2}{x}
\end{equation}
and $x_{0}=x_{0}(\varphi)>0$ satisfies
\begin{equation}
h'(x_0)=0\Leftrightarrow
x_0^3-x_0^2y\cos \varphi-1=0. 
\label{x0eq}
\end{equation}
Note that Eq.~\eqref{x0eq} has only one positive root 
as a function of $\varphi$. 

In Eq.~\eqref{geointx}, we again perform 
the stationary phase approximation in the integral 
with respect to $\varphi$, 
and we obtain 
\begin{eqnarray}
F\simeq F_{\rm geo}&=&
\Biggl(
\frac{x_+^3}{\sqrt{(x_+^3+2)(x_+^3-1)}}\exp\left[i\omega d\frac{-x_+^3+4}{x_+}\right]\cr
&&+
\frac{x_-^3}{\sqrt{(x_-^3+2)(1-x_-^3)}}\exp\left[i\omega d\frac{-x_-^3+4}{x_-}
-\frac{i\pi}{2}\right]\Biggr), 
\end{eqnarray}
where $x_\pm$ satisfies 
\begin{equation}
x_\pm^3\mp x_\pm^2y-1=0. 
\end{equation}
Note that $1<x_+$ and $0<x_-<1$. 
If we define $\mu_\pm$ and $\theta_\pm$ as 
\begin{eqnarray}
\mu_\pm&=&\frac{x_\pm^6}{(x_\pm^3+2)(x_\pm^3-1)}, \\
\theta_\pm&=&\omega d\frac{-x_\pm^3+4}{x_\pm}-\frac{\pi}{4}\pm\frac{\pi}{4}, 
\end{eqnarray}
$F_{\rm geo}$ can be expressed as 
\begin{equation}
F_{\rm geo}=\sum_\pm\sqrt{|\mu_\pm|}\ee^{i\theta_\pm}. 
\end{equation}
$\mu_\pm$ is the magnification factor 
for each image in the geometrical optics approximation. 

\begin{figure}[htbp]
\includegraphics[scale=0.6]{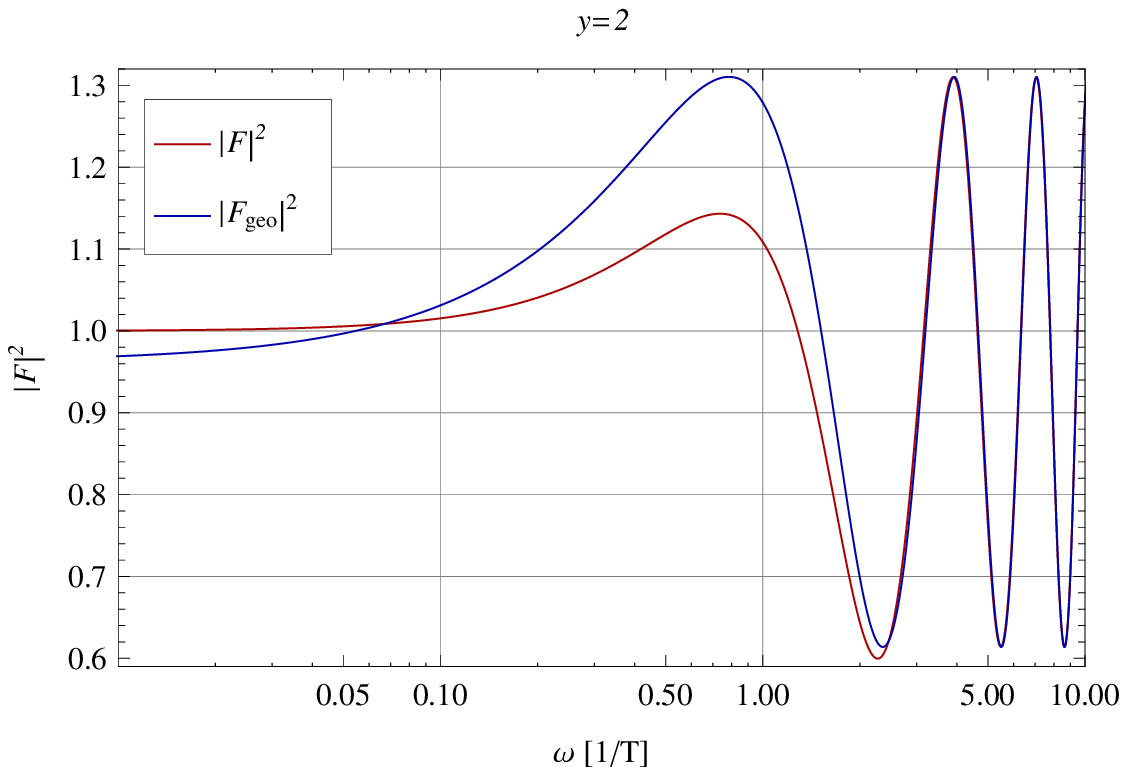}
\includegraphics[scale=0.6]{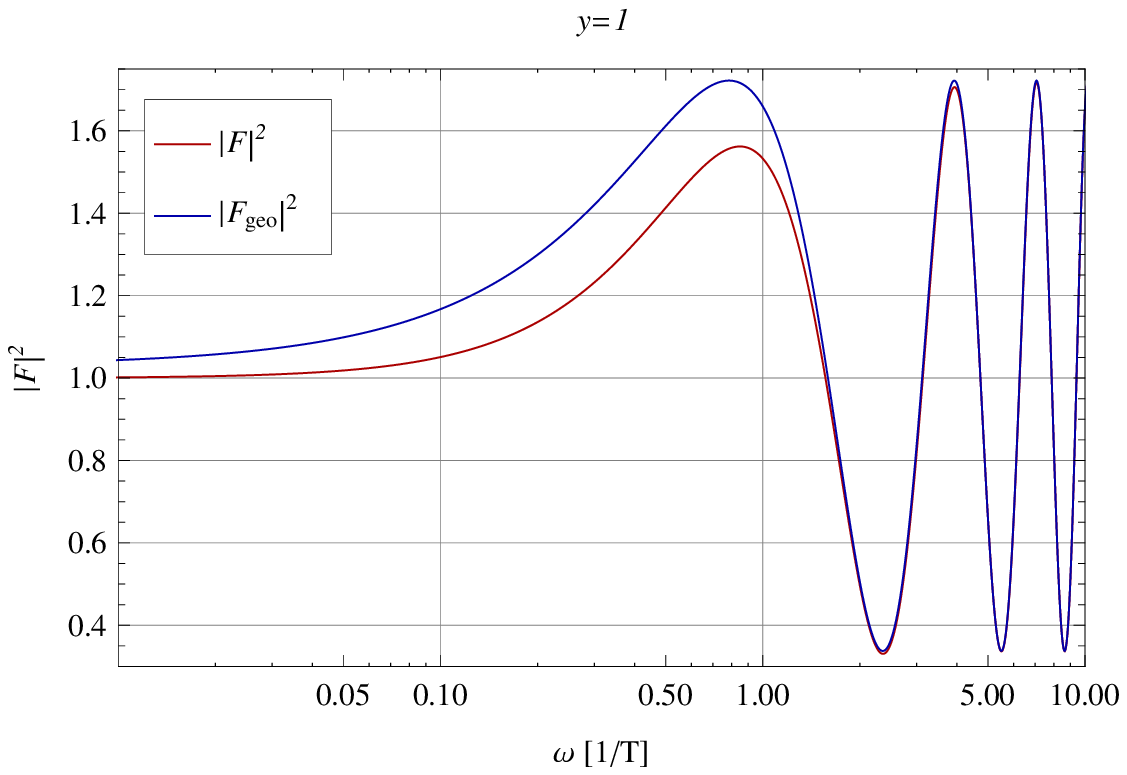}

\includegraphics[scale=0.6]{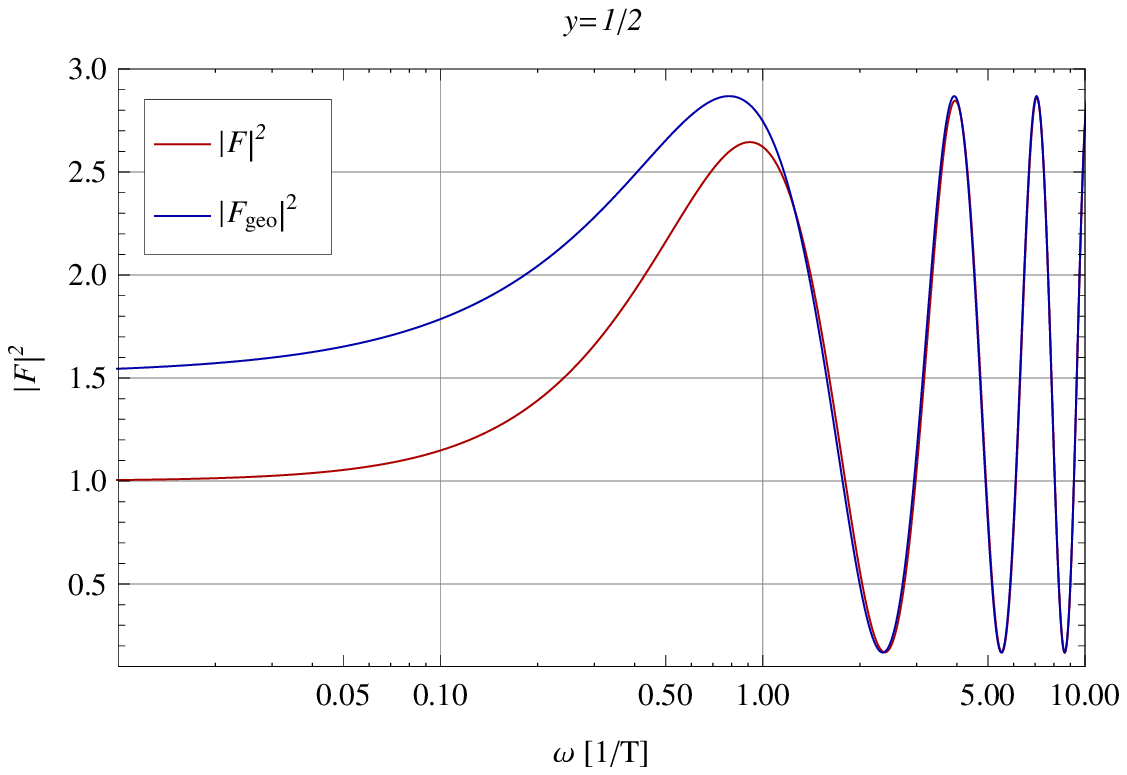}
\includegraphics[scale=0.6]{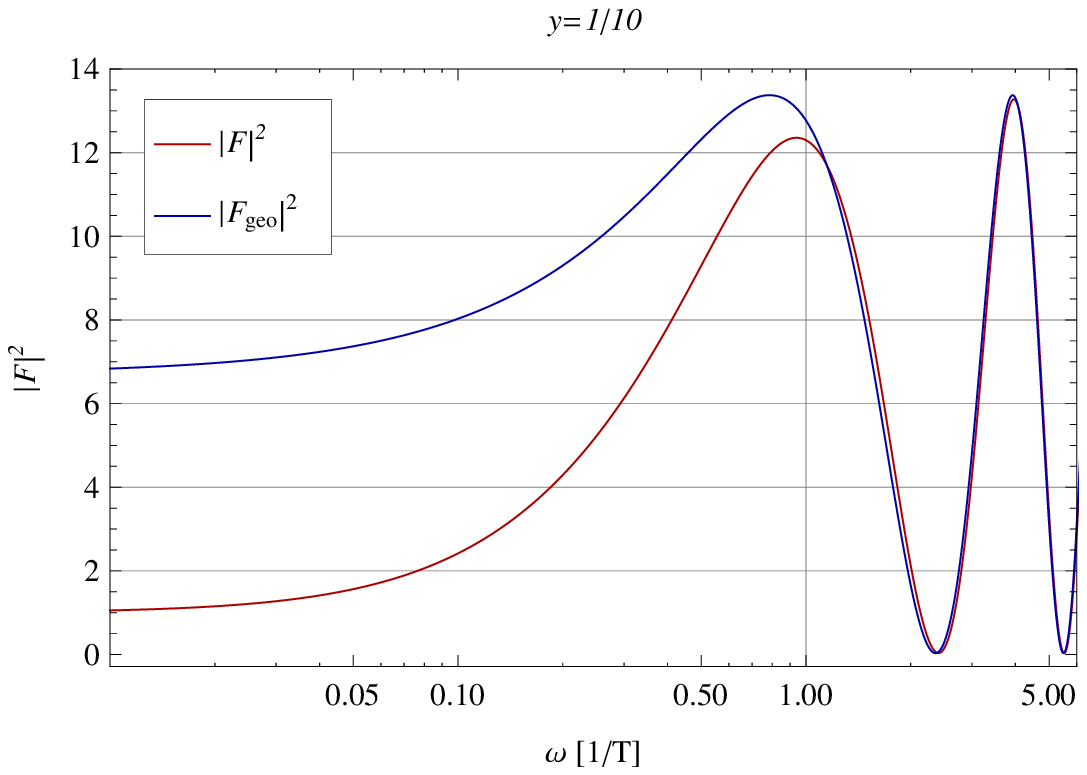}
\caption{$|F|^2$ and $|F_{\rm geo}|^2$ for the wormhole lens, where $T=d\,\tau $.  
}
\label{waveformwh}
\end{figure}

As an observable, we focus on $|F|^2$ in this paper. 
In the geometrical optics approximation, we obtain
\begin{equation}
|F_{\rm geo}|^2=|\mu_+|+|\mu_-|+2\sqrt{|\mu_+\mu_-|}\sin(2\omega d \tau(y)) 
\label{maggeo}
\end{equation}
with $\tau(y)$ being the following:
\begin{equation}
\tau(y):=\frac{\theta_--\theta_++\pi/2}{2\omega d}, 
\end{equation}
where note that the definition of $\tau$ can be written in terms of 
$x_\pm$, which is a function of $y$. 
$|F|^2$ and $|F_{\rm geo}|^2$ are depicted as functions of $\omega$ 
for each value of $y$ in Fig.~\ref{waveformwh}.
%%%%%%%%%%%%%%%%%%%%%%%%%%%<<start figure>>%%%%%%%%%%%%%%%%%%%%%%%%%%
%%%%%%%%%%%%%%%%%%%%%%%%%%%%<<end figure>>%%%%%%%%%%%%%%%%%%%%%%%%%%%
%

%%%%%%%%%%%%%%%%%%%%%%%%%%%%%%%%%%%%%%%%%%%%%%%%%%%%%%%%%%%%%%%%
\section{Comparison with the point mass lens based on observables}
\label{obs}
%%%%%%%%%%%%%%%%%%%%%%%%%%%%%%%%%%%%%%%%%%%%%%%%%%%%%%%%%%%%%%%%

In this paper, we assume the following situations for the observation:
\begin{itemize}
\item We can observe the spectrum of a source. 
\item The unlensed 
spectrum shape is well known. 
\end{itemize}
Note that, in our analysis, 
knowledge about the luminosity is not necessary. 
The amplification factor for the point mass lens is 
summarized in Appendix~\ref{pml}. 
For both point mass and wormhole cases, in the geometrical optics approximation, 
we obtain the form of Eq.~\eqref{maggeo} or equivalently Eq.~\eqref{eq:pmgeo}.

\begin{figure}[htbp]
\includegraphics[scale=]{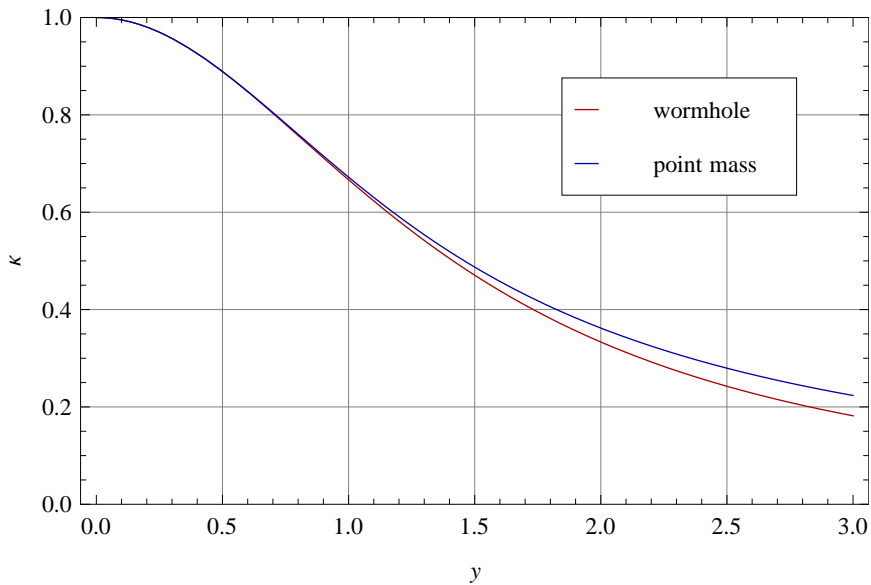}
\caption{$\kappa$ as a function of $y$ for the Ellis wormhole case and 
the point mass lens case. 
}
\label{ratio}
\end{figure}

\begin{figure}[htbp]
\includegraphics[scale=]{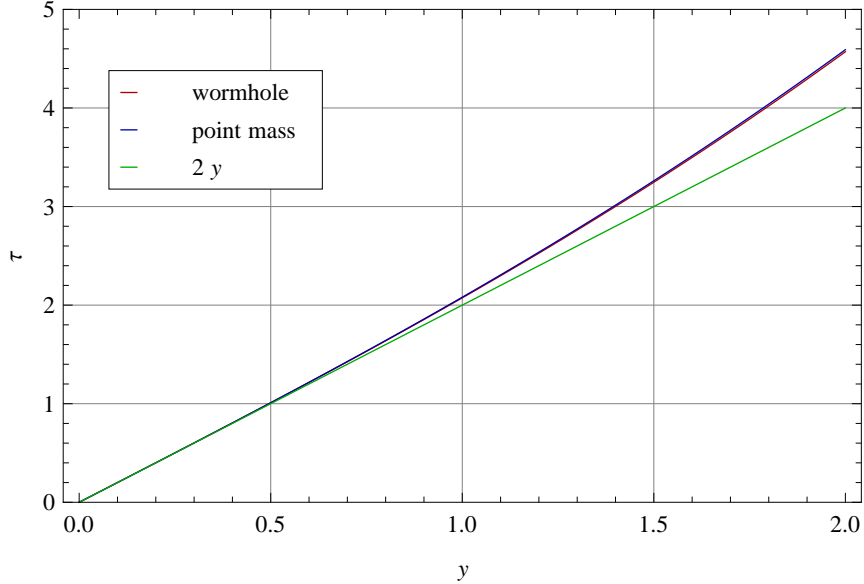}
\caption{$\tau(y)$ for the Ellis wormhole case and the point mass lens case. 
The two cases are almost %coincide with 
indistinguishable from each other in 
the region depicted here. 
}
\label{isou}
\end{figure}

In the frequency region where the geometrical optics approximation 
is valid, 
there are basically three observables 
which characterize the form of the amplification factor. 
The first 
is the frequency $\omega$,  
the second is the period of the oscillation of 
the spectrum as a function of $\omega$, and the third is the ratio 
$\kappa$ between the amplitude of the oscillation and 
the mean value. 
The period of the oscillation of the spectrum 
makes $T:=\tau(y)d$ an observable. 
$\kappa$ is given by
\begin{equation}
\kappa=\frac{2\sqrt{|\mu_+\mu_-|}}{|\mu_+|+|\mu_-|} 
\end{equation}
from Eq.~\eqref{maggeo} and 
%$\kappa$ can be 
plotted as a function of $y$ 
as is shown in Fig.~\ref{ratio}. 
%
%%%%%%%%%%%%%%%%%%%%%%%%%%%<<start figure>>%%%%%%%%%%%%%%%%%%%%%%%%%%
%%%%%%%%%%%%%%%%%%%%%%%%%%%%<<end figure>>%%%%%%%%%%%%%%%%%%%%%%%%%%%
Since $\kappa$ is observable, $y$ and hence $\tau(y)$ can be 
determined 
if we have enough accuracy of the observation. 
As shown in Fig.~\ref{isou}, 
$\tau(y)$ is a monotonically increasing  
function of $y$ and 
close to $2y$ in the region $y<1$. 
%%%%%%%%%%%%%%%%%%%%%%%%%%%<<start figure>>%%%%%%%%%%%%%%%%%%%%%%%%%%
%%%%%%%%%%%%%%%%%%%%%%%%%%%%<<end figure>>%%%%%%%%%%%%%%%%%%%%%%%%%%%
%
Then, from another observable $T=\tau(y) d$ we can obtain the 
value of $d$. 

\begin{figure}[htbp]
\includegraphics[scale=0.6]{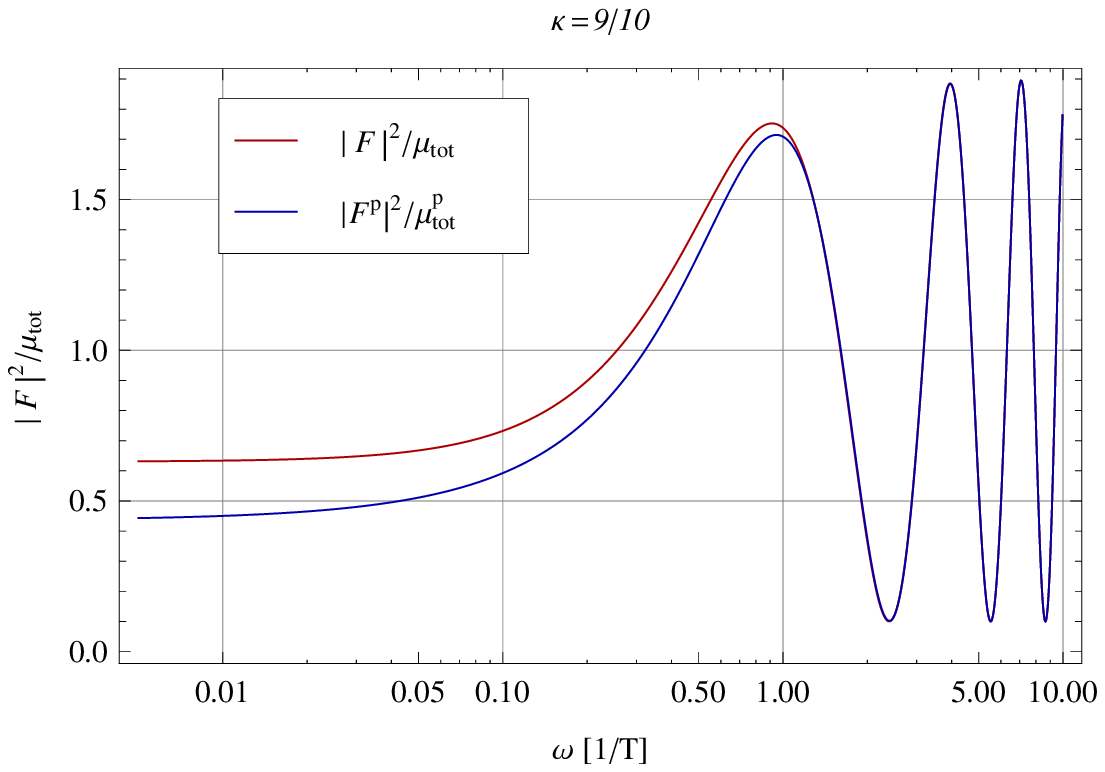}
\includegraphics[scale=0.6]{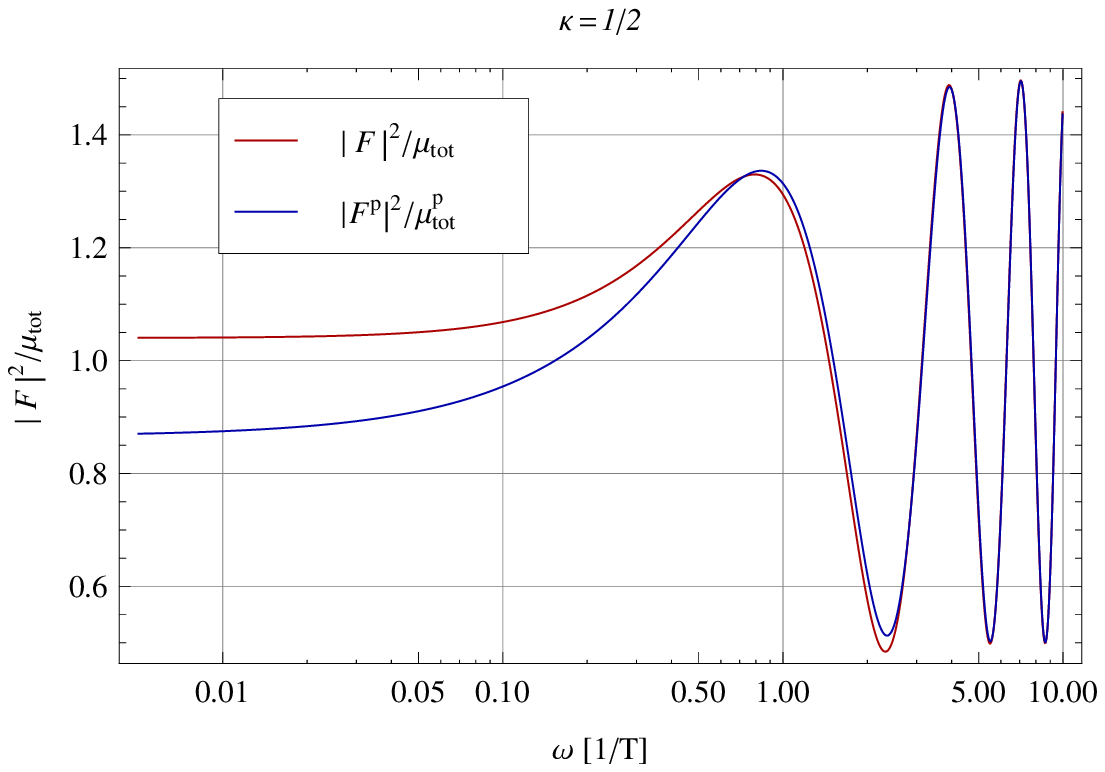}
\caption{Comparison between $|F|^2$ and $|F^{\rm p}|^2$. 
}
\label{compare}
\end{figure}

The situation for  
the point mass lens case 
is the same as for the Ellis wormhole case. 
That is, the three observables $\omega$, $T$, and $\kappa$ 
can be %explained by using 
regarded as %the 
gravitational lensing by a %the 
point mass as well as 
%the 
a wormhole. 
This fact indicates that 
we cannot %identify 
distinguish which is the lens object only
by using these three observables in the geometrical 
optics approximation. 
This degeneracy is resolved in 
the small frequency region in which the wave effect becomes significant as is explicitly shown in Fig.~\ref{compare}. 
%%%%%%%%%%%%%%%%%%%%%%%%%%%<<start figure>>%%%%%%%%%%%%%%%%%%%%%%%%%%
%%%%%%%%%%%%%%%%%%%%%%%%%%%%<<end figure>>%%%%%%%%%%%%%%%%%%%%%%%%%%%

\begin{figure}[htbp]
\includegraphics[scale=1]{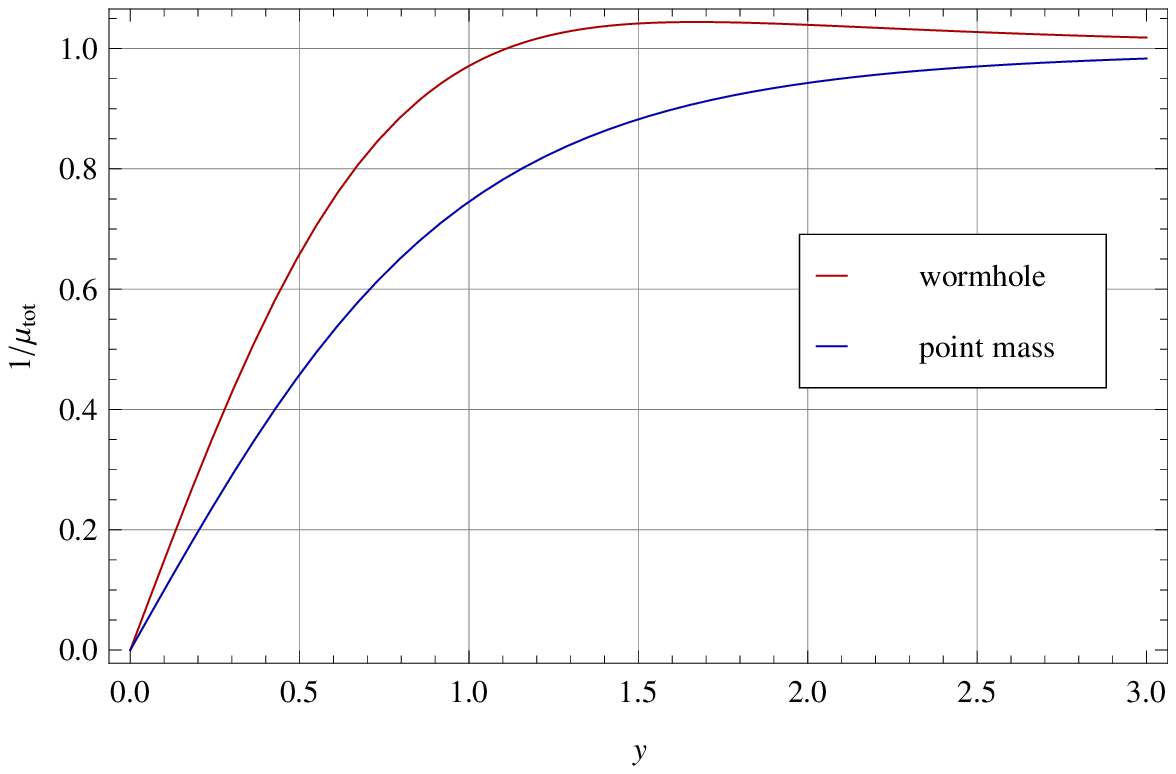}
\caption{$1/\mu_{\rm tot}$
}
\label{magratio}
\end{figure}

To simply see this resolution of the degeneracy, 
we consider the small frequency limit, 
i.e., $\omega\rightarrow0$. 
In this limit, we have $F\rightarrow 1$ as explicitly shown in Figs.~\ref{waveformwh} 
and \ref{waveformpo}. 
If we can observe the spectrum in any frequency region of interest, the
following quantity is an observable: 
\begin{equation}
\frac{\displaystyle \lim_{\omega\to 0}|F|^2}{\displaystyle \lim_{\omega\to \infty}<|F|^2>}
=1/\mu_{\rm tot}:=1/(|\mu_+|+|\mu_-|), 
\end{equation}
where the bracket $<~>$ denotes the average through several periods. 
We depict $1/\mu_{\rm tot}$ as a function of $y$ for both the wormhole case and the point mass lens case in Fig.~\ref{magratio}. 
%%%%%%%%%%%%%%%%%%%%%%%%%%%<<start figure>>%%%%%%%%%%%%%%%%%%%%%%%%%%
%%%%%%%%%%%%%%%%%%%%%%%%%%%%<<end figure>>%%%%%%%%%%%%%%%%%%%%%%%%%%%
In both cases, $1/\mu_{\rm tot}$ approaches to 0 and 1 in the limits 
$y\to 0$ and $y\to \infty$, respectively. 
While $1/\mu_{\rm tot}<1$ is satisfied in all domain of $y$ for the point mass lens, 
$1/\mu_{\rm tot}$ can exceed 1 for the wormhole case due to 
the demagnification effect originated from the negative mass density 
surrounding the Ellis wormhole. 
The behaviours of $1/\mu_{\rm tot}$ are totally different from each other. 
This fact shows that they are, in principle, distinguishable.

%%%%%%%%%%%%%%%%%%%%%%%%%%%%%%%%%%%%%%%%%%%%%%%%%%%%%%%%%%%%%%%%
\section{Observational constraint}
\label{obsconst}
%%%%%%%%%%%%%%%%%%%%%%%%%%%%%%%%%%%%%%%%%%%%%%%%%%%%%%%%%%%%%%%%
One of the possible sources is 
gamma-ray bursts, which have  
been proposed to be used for probing small mass primordial black holes\cite{1992ApJ...386L...5G,1993ApJ...413L...7S,Ulmer:1994ij,Marani:1998sh} and 
low tension cosmic strings\cite{Yoo:2012dn}. 
Recently, the femto-lensing effects caused by 
compact objects were searched by using gamma-ray bursts with known redshifts 
detected by the Fermi Gamma-ray Burst Monitor\cite{Barnacka:2012bm}. 
From non-detection of the femto-lensing event, 
a constraint on the number density of compact objects 
have been obtained. 
For a fixed value of the mass $M$ of each compact object, 
the constraint can be translated into a constraint on $\Omega_{\rm CO}$,
where $\Omega_{\rm CO}$ is the average energy density of 
compact objects in the unit 
of the critical density $\rho_{\rm cr}$. 
Then, the abundance of dark compact objects 
is constrained as $\Omega_{\rm CO}<0.15$ at the 
95\% confidence level for $M\sim 3\times10^{18}{\rm g}$.

Since the number density $n$ of the compact objects is given by 
\begin{equation}
n=\frac{\Omega_{\rm CO}\rho_{\rm cr}}{M}
\sim 2\times 10^{-9}{\rm AU}^{-3}
\left(\frac{\Omega_{\rm CO}}{0.15}\right)
\left(\frac{M}{3\times 10^{18}{\rm g}}\right)^{-1}
\left(\frac{\rho_{\rm cr}}{10^{-29}{\rm g/cm^3}}\right), 
\end{equation}
we obtain the constraint for the number dnsity 
of compact objects with $M\sim 3\times10^{18}{\rm g}$ 
as $n<2\times 10^{-9}{\rm AU}^{-3}$. 
Then, we can expect a similar constraint on the number density of 
Ellis wormholes with the throat radius which gives the same value of $d$ 
as that for the compact object. 
The mass of $3\times 10^{18}{\rm g}$ gives $d\sim 5\times10^{-10}\cm$ 
from Eq.~\eqref{dy4po} 
and, from Eq.~\eqref{d4wh}, 
the corresponding throat radius $a$ is given by 
\begin{equation}
a\sim 0.7 \cm\left(\frac{d}{5\times 10^{-10}\cm}\right)^{3/4}
\left(\frac{D}{10^{28}\cm}\right)^{1/4}. 
\end{equation}
Therefore, the number density of Ellis wormholes with $a\sim 1\cm$ 
must satisfy $n\lesssim 10^{-9}{\rm AU}^{-3}$.
Note that this constraint 
comes from the wave form in the geometrical optics approximation
and hence we do not distinguish between the point mass lenses and Ellis 
wormholes.

Another possible observation to probe Ellis wormholes is 
the observation of gravitational waves from compact object binaries. 
The unlensed 
wave form of the gravitational waves from a compact object binary
is well known. 
From the chirp signal in the inspiral phase, 
we can obtain the spectrum of the gravitational waves. 
In order to distinguish the Ellis wormhole from 
a point mass lens, 
we need to observe not only the typical interference pattern 
but also the wave effect in the lensed spectrum.  
Hence, the spectrum in $d\sim \lambda:=2\pi/\omega$ 
is necessary to probe the Ellis wormhole. 
Assuming $d\sim \lambda$, 
the typical throat radius of Ellis wormholes which can be 
probed by using gravitational waves is 
estimated as follows:
\begin{eqnarray}
a=d^{3/4}\left(\frac{2D}{\pi^2}\right)^{1/4}
&\sim&\lambda^{3/4}
\left(\frac{2D}{\pi^2}\right)^{1/4}\cr
&\sim&7\times 10^{12} \cm
\left(\frac{\lambda}{10^8\cm}\right)^{3/4}
\left(\frac{D}{10^{28}\cm}\right)^{1/4}. 
\end{eqnarray}

The same estimate is applicable for 
galactic sources of electro-magnetic waves. 
We obtain $a\sim10^5\cm$ for galactic radio sources 
with $\lambda \sim 1\cm$, 
$a\sim 1$m for galactic optical or infra-red sources
and $a\sim 1$cm for galactic X-ray sources. 
The source must be compact enough to show  
the clear oscillation behaviour in the spectrum. 
This fact can be clearly understood by 
considering the $y$ dependence of the phase in 
the amplification factor \eqref{maggeo}. 
Since $\tau(y)$ is roughly approximated by $2y$, 
the period $\delta y$ for one cycle is 
given by $\delta y\sim \pi/(\omega d)$. 
The corresponding length scale $\delta \eta$ 
on the source plane is given by 
\begin{eqnarray}
\delta \eta
=\delta y \frac{\ds}{\dl}\xi_0
&\sim&\frac{\pi}{\omega}\sqrt{\frac{2\dls \ds}{d\dl}}\cr
&\sim&\sqrt{\frac{\lambda \dls\ds}{2\dl}}
\sim2\times 10^{11}\cm \left(\frac{\lambda}{1\cm}\right)^{1/2}
\left(\frac{\dls\ds/\dl}{10 {\rm kpc}}\right)^{1/2}, 
\end{eqnarray}
where we have assumed $d\sim\lambda$. 
If the source radius is larger than $\delta \eta$, 
the interference pattern will be smeared out. 
Observation of compact galactic sources such as pulsars and white dwarfs might 
be useful to probe not only dark compact objects but also exotic compact objects 
such as the Ellis wormhole. 

%%%%%%%%%%%%%%%%%%%%%%%%%%%%%%%%%%%%%%%%%%%%%%%%%%%%%%%%%%%%%%%%
\section{summary}
\label{summary}
%%%%%%%%%%%%%%%%%%%%%%%%%%%%%%%%%%%%%%%%%%%%%%%%%%%%%%%%%%%%%%%%

In this paper, we have proposed 
the probe of Ellis wormholes by using spectroscopic observations. 
We have assumed that 
the spectrum of the target source can be measured in 
enough accuracy and the spectrum shape is well known 
without lensing, but the luminosity is not necessarily observable. 
Then, we have discussed the distinguishability of the 
lensed spectrum from the case of the point mass lens. 

We have derived the wave form 
after the scattering by the Ellis wormhole 
including the wave effect in the low frequency domain. 
The geometrical optics limit of the wave form has been also 
analytically derived. 
Then, we have found that the Ellis wormhole cannot be distinguished from 
the point mass lens 
by using only the high frequency domain in which the geometrical optics approximation 
is valid. 
We have also found that this degeneracy is resolved in 
the low frequency domain in which the wave effect is significant. 
Possible observational constraints are also 
discussed and 
we estimated the upper bound for the 
number density of Ellis wormholes as 
$n\lesssim 10^{-9}{\rm AU}^{-3}$ 
with throat radius $a\sim 1\cm$ from 
the existing femto-lensing analysis for 
compact objects. 

Finally, we note that our method to probe the Ellis wormhole 
is complementary to the other methods to 
probe the Ellis wormhole with micro-lensing \cite{Safonova:2001vz,Bogdanov:2008zy,Abe:2010ap} or the astrometric image
centroid displacements\cite{Toki:2011zu}. 
These are 
not feasible for observations on cosmological scales 
because the time scale of the lens event is too long 
to detect modulation of the light curve or the displacements.
In contrast, the slow relative motion is 
an advantage for spectroscopic observations. 
Therefore we may probe the Ellis wormhole on cosmological scales 
using our method.

\section*{Acknowledgements}
CY is supported by a Grant-in-Aid through the
Japan Society for the Promotion of Science (JSPS).
The work of NT was supported in part by Rikkyo University
Special Fund for Research. 
TH was supported by the Grant-in-Aid for Young Scientists (B) (No. 21740190) 
and the Grant-in-Aid for Challenging Exploratory Research (No. 23654082) 
for Scientific Research Fund of the Ministry of Education, Culture, Sports, Science and Technology, Japan.

\appendix

%%%%%%%%%%%%%%%%%%%%%%%%%%%%%%%%%%%%%%%%%%%%%%%%%%%%%%%%%%%%%%%%
\section{Derivation of Eq.~\eqref{sourceint3}}
\label{dereq}
%%%%%%%%%%%%%%%%%%%%%%%%%%%%%%%%%%%%%%%%%%%%%%%%%%%%%%%%%%%%%%%%
The first term in Eq.~\eqref{sourceint2} 
can be evaluated as follows. 
\begin{eqnarray}
\int^{\rm B}_{\rm S} dl=
|\overrightarrow{\rm SA}|+|\overrightarrow{\rm AB}|
&=&|\overrightarrow{\rm SA}|+|\overrightarrow{\rm OA}-\overrightarrow{\rm OB}|\cr
&=&|\overrightarrow{\rm SA}|
+\sqrt{|\overrightarrow{\rm OA}|^2+|\overrightarrow{\rm OB}|^2
-2\overrightarrow{\rm OA}\cdot\overrightarrow{\rm OB}}. 
\end{eqnarray}
$\overrightarrow{\rm OB}$ and $\overrightarrow{\rm OB'}$ can be written as 
\begin{eqnarray}
\overrightarrow{\rm OB}&=&
\left[\left(1-\frac{\delta D}{\dl}\right)
\overrightarrow{\rm OL}
+\vec \eta+\left(\vec \xi-\vec \eta\right)\frac{\dls+\delta D}{\dls}
-\alpha\delta D\frac{\vec \xi}{\xi}\right](1+\mathcal O(\epsilon^3))\cr
&=&\left[\left(1-\frac{\delta D}{\dl}\right)
\overrightarrow{\rm OL}
+\vec \xi -\alpha(\xi)\delta D \frac{\vec \xi}{\xi}
+\left(\vec \xi-\vec \eta\right)\frac{\delta D}{\dls}\right](1+\mathcal O(\epsilon^3)),\\
\overrightarrow{\rm OB'}&=&\left(1-\frac{\delta D}{\dl}\right)\overrightarrow{\rm OA}. 
\end{eqnarray}
From these expressions, we can find 
\begin{eqnarray}
|\overrightarrow{\rm OB}|&=&
|\overrightarrow{\rm OB'}|\left(1+\mathcal O(\epsilon^3)\right), \\
\overrightarrow{\rm OA} \cdot \overrightarrow{\rm OB} 
&=&\overrightarrow{\rm OA} \cdot \overrightarrow{\rm OB'}
\left(1+\mathcal O(\epsilon^3)\right). 
\end{eqnarray}
Therefore we obtain 
\begin{equation}
\int^{\rm B}_{\rm S} dl=
\left(|\overrightarrow{\rm SA}|+|\overrightarrow{\rm OA}| 
-|\overrightarrow{\rm OB'}|\right)\left(1+\mathcal O(\epsilon^3)\right). 
\end{equation}
Since we find
\begin{eqnarray}
|\overrightarrow{\rm SA}|
&=&\sqrt{|\vec \xi-\vec \eta|^2+\dls^2}
=\dls\left(1+\frac{|\vec \xi-\vec \eta|^2}{2\dls^2}+\mathcal O(\epsilon^4)\right), \\
|\overrightarrow{\rm AO}|&=&
\dl\left(1+\frac{\xi^2}{2\dl^2}+\mathcal O(\epsilon^4)\right), 
\end{eqnarray}
we obtain the following expression:
\begin{equation}
\int^{\rm B}_{\rm S} dl=
\left[
\ds\left(1+\frac{\eta^2}{2\ds^2}\right)+
\frac{\dl\ds}{2\dls}\left(\frac{\vec \xi}{\dl}-\frac{\vec \eta}{\ds}\right)^2
-r\right]\left(1+\mathcal O(\epsilon^3)\right), 
\end{equation}
where $r=|\overrightarrow{\rm OB'}|$. 

In order to evaluate the second term in Eq.~\eqref{sourceint2},
we first consider the integral between S and A. 
Letting P be a point on the segment SA, 
we obtain 
\begin{equation}
|\overrightarrow{\rm LP}|^2=|\vec \xi+\overrightarrow{\rm AP}|^2
=\left|\vec \xi+\left(1-\frac{l}{l_{\rm SA}}\right) \overrightarrow{\rm AS}\right|^2
=\xi^2+\left(l_{\rm SA}-l\right)^2+2\left(1-\frac{l}{l_{\rm SA}}\right)\vec \xi\cdot 
\overrightarrow{\rm AS}, 
\end{equation}
where $l_{\rm SA}=|\overrightarrow{\rm SA}|$ and $l=|\overrightarrow{\rm SP}|$. 
Since $\vec \xi\cdot \overrightarrow{\rm AS}=-\vec \xi\cdot (\vec \xi-\vec \eta)$, 
we obtain the following expression:
\begin{equation}
|\overrightarrow{\rm LP}|^2
=\xi^2+\left(l_{\rm SA}-l\right)^2-2\left(1-\frac{l}{l_{\rm SA}}\right)
\vec \xi \cdot (\vec \xi-\vec \eta). 
\end{equation}
Substituting the above expression of $|\overrightarrow{\rm LP}|^2$ 
into $R^2$ of the second integral in Eq.~\eqref{sourceint2} 
with the integral region being from S to A, 
we obtain 
\begin{eqnarray}
\int^{\rm A}_{\rm S} \frac{1}{R^2} dl=\int^{l_{\rm SA}}_{0}
\frac{dl}{\xi^2+(l_{\rm SA}-l)^2-2(1-\frac{l}{l_{\rm SA}})
\vec \xi \cdot (\vec \xi-\vec \eta)}. 
\label{secondint1}
\end{eqnarray}
Then the integral \eqref{secondint1} can be performed and evaluated as 
\begin{eqnarray}
\int^{l_{\rm SA}}_{0}
\frac{dl}{(l_{\rm SA}-l)^2+\xi^2-2(1-\frac{l}{l_{\rm SA}})
\vec \xi \cdot (\vec \xi-\vec \eta)}
&=&\frac{l_{\rm SA}}{\xi \dls}\left[\arctan\left(
\frac{-l_{\rm SA}^2+\vec \xi \cdot (\vec \xi-\vec \eta)
+l l_{\rm SA}}{\xi \dls}\right)\right]^{l_{\rm SA}}_0\cr
&=&\left(\frac{\pi}{2\xi}-\frac{\vec\xi\cdot\vec\eta}{\xi^2 \dls}\right)\left(1+\mathcal O(\epsilon^2)\right). 
\end{eqnarray}
The contribution from the integral between A and B can be also evaluated 
by the similar integral. 
Finally, we obtain the expression \eqref{sourceint3}.

%%%%%%%%%%%%%%%%%%%%%%%%%%%%%%%%%%%%%%%%%%%%%%%%%%%%%%%%%%%%%%%%
\section{Point Mass Lens}
\label{pml}
%%%%%%%%%%%%%%%%%%%%%%%%%%%%%%%%%%%%%%%%%%%%%%%%%%%%%%%%%%%%%%%%
For the point mass case, we have the following expression for 
the amplification factor~\cite{Deguchi:1986zz,1992grle.book.....S}:
\begin{equation}
\left|F^{\rm po}\right|=\left|\ee^{\pi \omega d/2}\Gamma\left(1-i\omega d\right)
{}_1F_1\left(i\omega d,1;i\omega dy\right)\right|, 
\end{equation}
where $\Gamma$ and ${}_1F_1$ are the gamma function and the 
confluent hyper-geometric function, 
respectively, and 
\begin{equation}
d=2M,
\quad 
y=\eta\sqrt{\frac{D_{\rm L}}{4MD_{\rm LS}D_{\rm S}}}. 
\label{dy4po}
\end{equation}

\begin{figure}[htbp]
\includegraphics[scale=0.6]{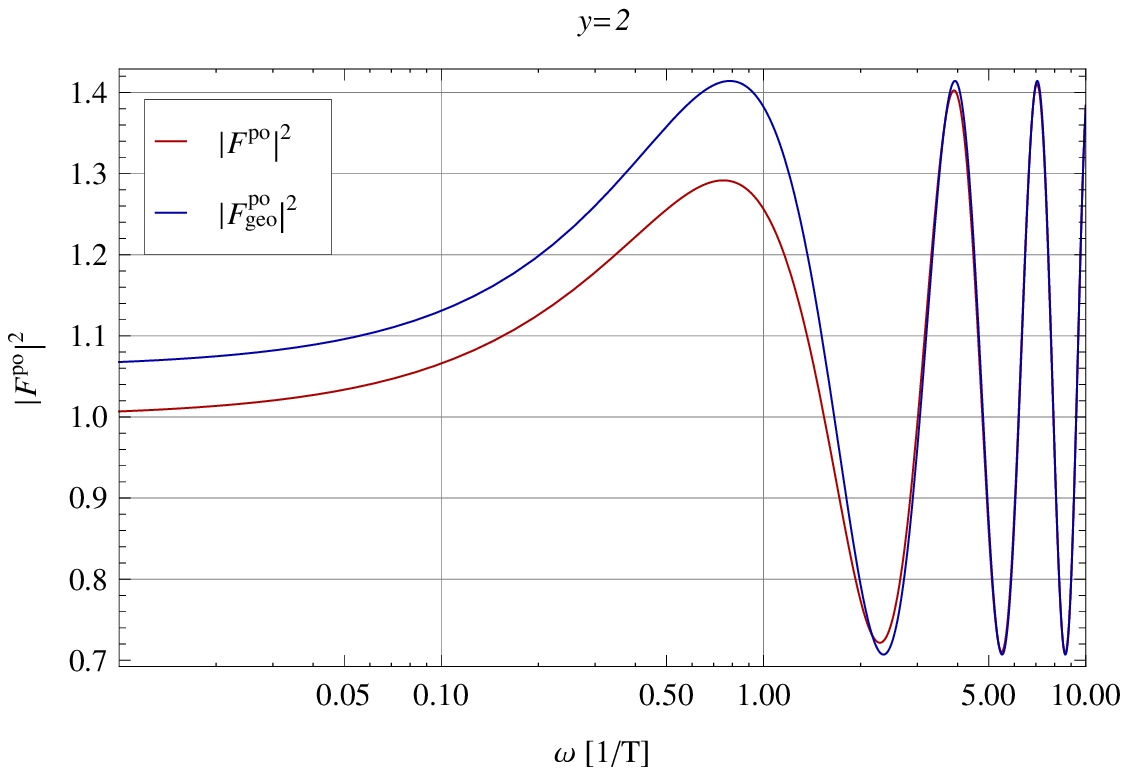}
\includegraphics[scale=0.6]{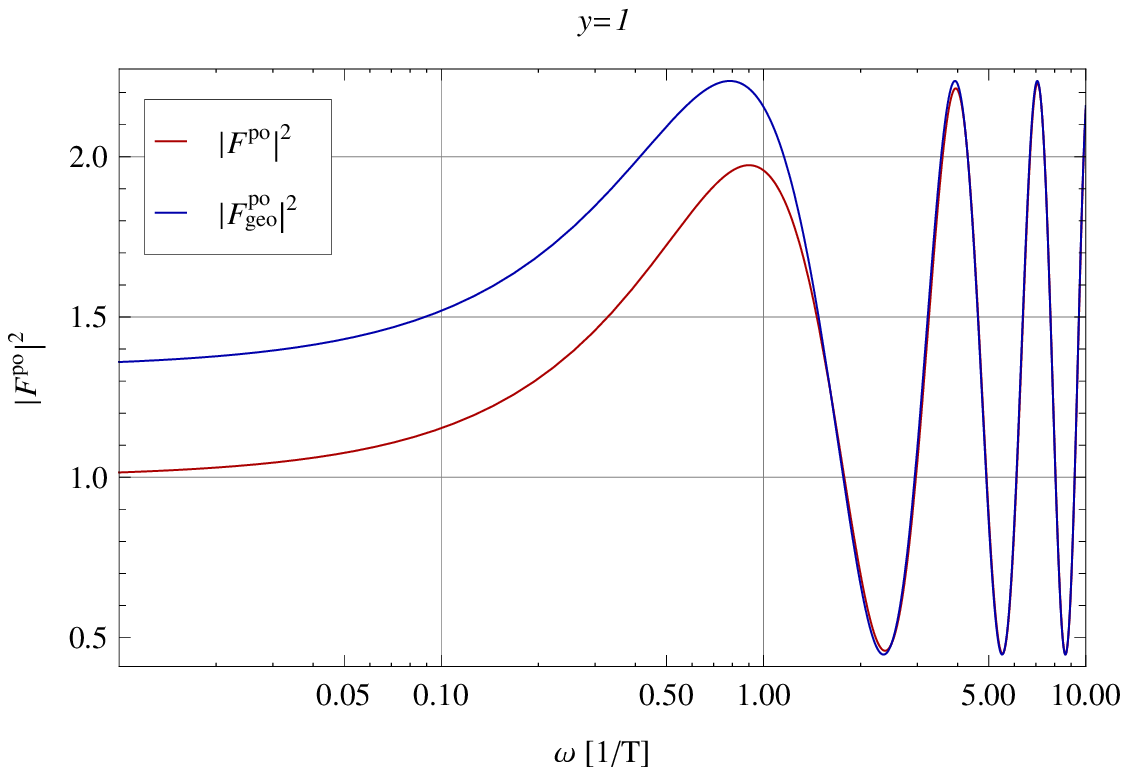}

\includegraphics[scale=0.6]{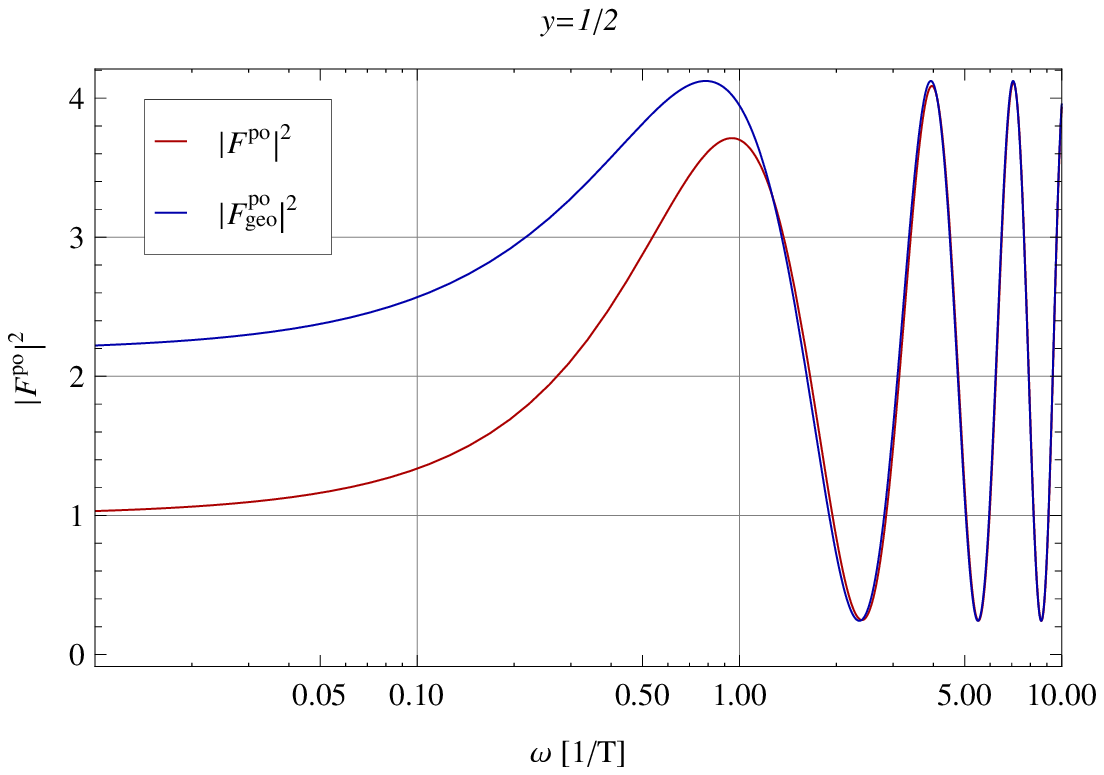}
\includegraphics[scale=0.6]{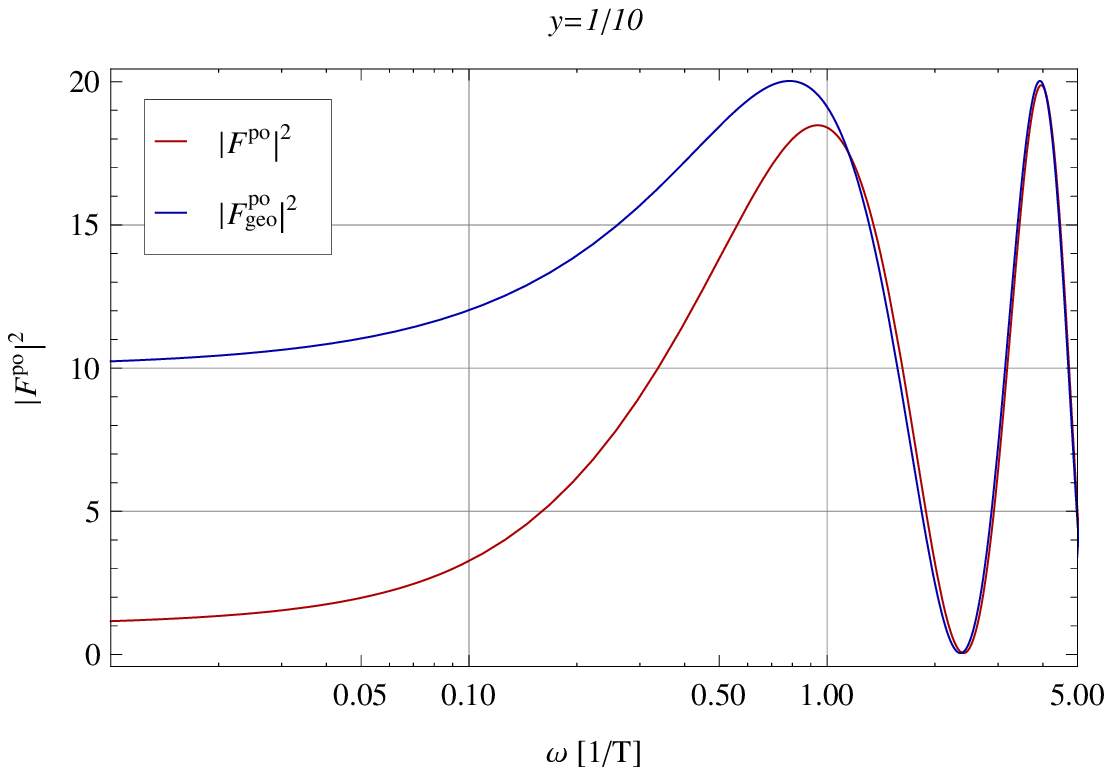}
\caption{
$|F^{\rm po}|^2$ and $|F^{\rm po}_{\rm geo}|^2$ as 
functions of $\omega$ for each value of $y$, 
where $T=d\,\tau_{\rm po}$. 
}
\label{waveformpo}
\end{figure}

In the geometrical optics approximation ($\omega\rightarrow \infty$), 
we obtain 
\begin{equation}
\left|F^{\rm po}\right|^2\rightarrow\left|F^{\rm po}_{\rm geo}\right|^2
:=\left|\mu^{\rm po}_+\right|+\left|\mu^{\rm po}_-\right|
+2\sqrt{\left|\mu^{\rm po}_+\mu^{\rm po}_-\right|}\sin(2\omega d \tau_{\rm po}(y)),
\label{eq:pmgeo}
\end{equation}
where
\begin{eqnarray}
\mu^{\rm po}_\pm&=&\pm \frac{1}{4}\left[\frac{y}{\sqrt{y^2+4}}
+\frac{\sqrt{y^2+4}}{y}\pm2\right], \\
\tau_{\rm po}(y)&=&\frac{1}{2}y\sqrt{y^2+4}
+\ln\frac{\sqrt{y^2+4}+y}{\sqrt{y^2+4}-y}. 
\end{eqnarray}
$|F^{\rm po}|^2$ and $|F^{\rm po}_{\rm geo}|^2$ are depicted as 
functions of $\omega$ for each value of $y$ in Fig.~\ref{waveformpo}. 
%%%%%%%%%%%%%%%%%%%%%%%%%%%<<start figure>>%%%%%%%%%%%%%%%%%%%%%%%%%%
%%%%%%%%%%%%%%%%%%%%%%%%%%%%<<end figure>>%%%%%%%%%%%%%%%%%%%%%%%%%%%

%\bibliographystyle{h-physrev5-title}
%\bibliography{../bibfiles/whlens}

\end{document}